\documentclass[12pt,preprint]{aastex}

\begin{document}

\def\chandra{{\em Chandra}}

\title{Revealing New Physical Structures in the Supernova Remnant N63A 
through {\em Chandra} Imaging Spectroscopy}

\author{Jessica S. Warren, John P. Hughes}
\affil{Department of Physics and Astronomy, Rutgers University}
\affil{136 Frelinghuysen Road, Piscataway, NJ 08854-8019}
\email{jesawyer@physics.rutgers.edu, jph@physics.rutgers.edu}
\and
\author{Patrick O. Slane}
\affil{Harvard-Smithsonian Center for Astrophysics}
\affil{60 Garden Street, Cambridge, MA 02138}

\begin{abstract}
We present {\em Chandra} X-ray observations of the supernova remnant
(SNR) N63A in the Large Magellanic Cloud (LMC). N63A, one of the
brightest LMC remnants, is embedded in an H {\sc ii} region and
probably associated with an OB association. The optical remnant
consists of three lobes of emission contained within the approximately
three times larger X-ray remnant. Our \chandra\ data reveal a number
of new physical structures in N63A.  The most striking of these are
the several ``crescent''-shaped structures located beyond the main
shell that resemble similar features seen in the Vela SNR. In Vela,
these have been interpreted as arising from high speed clumps of
supernova ejecta interacting with the ambient medium. Another distinct
feature of the remnant is a roughly triangular ``hole'' in the X-ray
emission near the location of the optical lobes and the brightest
radio emission.  X-ray spectral analysis shows that this deficit of
emission is a result of absorption by an intervening dense cloud with
a mass of $\sim$450 $M_\odot$ that is currently being engulfed by the
remnant's blast wave. We also find that the rim of the remnant, as
well as the crescent-shaped features, have considerably softer X-ray
spectra than the interior.  Limits on hard X-ray emission rule out a
young, energetic pulsar in N63A, but the presence of an older or less
active one, powering a wind nebula with a luminosity less than
$\sim$$4\times 10^{34}$ erg s$^{-1}$, is allowed.

\end{abstract}

\section{INTRODUCTION}
Supernova remnants (SNRs) are valuable tools for understanding the
composition of the interstellar medium (ISM), recent star formation, and
supernovae (SNe) themselves.  The structure of SNRs and their
interactions with the surrounding medium give us insight into
their origin and effects on their environment.  The second brightest
X-ray SNR in the Large Magellanic Cloud (LMC), N63A, provides us with
an excellent laboratory for studying such structures and interactions.
This remnant is embedded in a larger H {\sc ii} region, N63, and
appears to be located within the OB association NGC 2030 \citep{chu2}.
N63A is believed to be the product of the explosion of a massive star
in a dense and complex environment \citep{shull,hughes} and is the
first confirmed SNR in an H {\sc ii} region \citep{shull}.

The X-ray size of the SNR ($34^{\prime\prime}$ or 8.2 pc radius for an
LMC distance of 50 kpc) is about three times the size of the optical
remnant, which appears as a three-lobed structure \citep{mathewson}.
Two lobes are shock-heated, while the third (southwestern) is a
photoionized H {\sc ii} region \citep{levenson}.  The shock-heating is
a result of interactions with the ISM, rather than with SN ejecta,
based on the derived abundances \citep{rusdop90}.  Similarly, the
X-ray emission from N63A is also consistent with swept-up ISM (Hughes
et al.~1998). In the radio, the remnant appears as a thick shell, with
the brightest emission corresponding to the two shock-heated eastern
lobes of optical emission \citep{dickel}.  N63A was shown by
\citet{graham} to be a significant IRAS source, with
$L_{\rm IR}=1.5\times10^5 L_\odot$.  However, the Columbia CO survey of
the LMC revealed no significant molecular gas associated with the SNR
\citep{cohen} and no CO emission was detected in a higher angular
resolution pointed observation by SEST \citep{israel}.  Hughes et
al.~(1998) estimate the age of N63A to be in the range 2000-5000 yr.

We targeted N63A with \chandra\ to study the interaction of the SNR
with the ISM, to find evidence for SN ejecta, and to search for a
compact remnant and its associated pulsar wind nebula (PWN).  Here we
report on the most prominent new features of the remnant derived from
the \chandra\ imaging and spectral data.

\section{OBSERVATIONS}

N63A was observed for 45.7 ks on 2000 October 16 on the
back-side--illuminated chip (S3) of the Advanced CCD Imaging
Spectrometer (ACIS-S) (Obs ID 777).  The remnant was offset slightly
from the aimpoint to ensure that it fell entirely on one node of the
chip.  In order to reduce as much of the pixel-to-pixel variation in
the data as possible, we employed the charge transfer inefficiency
(CTI) corrector code \citep{townsley} to correct for spatial gain
variations. We used standard software tools to filter for grade
(retaining grades 02346 only), bad pixels, and times of high or
flaring background.  There were count rate flares up to 26 s$^{-1}$
summed over the entire S3 chip (excluding the source), compared to a
rate of 24 s$^{-1}$ for N63A. Nevertheless the source extent is small
enough that even a moderate background count rate restriction ($<$14
s$^{-1}$) was found to be acceptable for our studies.  After all
filtering, the net exposure time was 41.4 ks.  All spectral analyses
utilized the PI column of the final {\em Chandra} events file.

In Figure 1a we present a three-color \chandra\ X-ray image of N63A.
This reverse-color image clearly shows the complex X-ray spatial and
spectral structure of this remnant.  Cyan (blue color) represents the
0.16--0.75 keV energy band; magenta (pink color) represents the
0.75--0.96 keV band, and yellow represents the 0.96--6.64 keV band.
These energy bands were chosen to obtain roughly equal numbers of
counts in the integrated \chandra\ spectrum of N63A.  Regions with
equal emission in each band should appear black in Figure 1a.  In the
southwestern and northern portions of the X-ray remnant
crescent-shaped features protrude far beyond the edge of the main
shell and appear bluish in color, indicating a relative enhancement in
soft emission.  A faint triangular region (henceforth, ``hole''), just
west of center, is characterized by higher energy emission.  The rest
of the interior of the remnant shows similar levels of emission in all
energy bands, although there is a broad greenish-yellow stripe that
runs diagonally from the northeast to the southwest.  Sharp linear
filaments extend across large portions of the remnant.  Relatively
bright knots and clumps of emission can be seen northwest of center
and in the east.  Images made in narrow X-ray energy bands reveal
subtle but significant variations.  The relative faintness of the
triangular region is most prominent in the lowest energy band but
becomes progressively less apparent toward higher energy.  At the
highest energies the remnant's emission fades slowly into the
background.  There is no hard unresolved X-ray source observed in
N63A.

The morphology of N63A varies strongly across the wavebands.  In
Figure 1b we present a color composite of the following: the Australia
Telescope Compact Array (ATCA) 8638 MHz image \citep{dickel} in red,
the {\em Hubble\/} Space Telescope WFPC2 H$\alpha$ image \citep{chu1}
in green, and the {\em Chandra} 0.4--7 keV broadband image in blue.
The brightest X-ray emission is toward the east where there is little
to no radio emission. Several compact optical knots evident in the
image do not correlate well with similar-sized structures in either
the X-ray or radio bands, although two appear near the brightest X-ray
emission.  The optical and brightest radio emission is broadly
associated with the triangular hole in the X-ray emission.  This is
discussed more in Section 4.

\section{ANALYSIS AND RESULTS}

\subsection{Image Analysis}

Our preliminary studies of the \chandra\ data suggested the presence
of significant column density variations with position in N63A. We
developed a simple technique to characterize the statistical
significance of these variations and to obtain numerical estimates of
the amount of absorption.  Initially we divided the entire remnant
into regions containing at least 1000 counts, using an adaptive mesh,
that resulted in 570 independent regions.  For each region a spectrum
was extracted over the 0.4--4.0 keV band and rebinned to a minimum of
25 events per channel. This was compared to the total observed
spectrum of the remnant, which was rebinned in the same way.

Two simple models were used for this comparison: one assumed that each
region's spectrum was simply a scaled-down version of the integrated
spectrum; the other allowed for both a scaling factor and an
additional X-ray absorption factor (hereafter $\Delta N_{\rm H}$). The
latter used abundances reduced to 0.3 times solar, as appropriate to
the LMC, to calculate $\Delta N_{\rm H}$.  Best-fit parameters in each
case were determined by minimizing $\chi^2$. We then compared these
two values with the $F$-test to determine the significance of the
fitted additional column density for each region.  We did not put any
\textit{a priori} limits on $\Delta N_{\rm H}$.

The most statistically significant $F$-values ($> 3\sigma$) correspond
to either highly negative or positive $\Delta N_{\rm H}$ values. A
negative $\Delta N_{\rm H}$ naively suggests a region with lower than
average absorbing line-of-sight column density, while a positive
$\Delta N_{\rm H}$ suggests a higher value. We plot these regions in
Figures 2b and 2c as soft and hard, respectively, where the grayscale
encodes the value of $\Delta N_{\rm H}$. Superposed on these images is
a slightly smoothed version of the broadband image, which is shown
(unsmoothed) in Figure 2a.  The regions of apparently negative $\Delta
N_{\rm H}$ are on the outer rim of the remnant, indicating spectral
softening at the shell. We plot an example spectrum of such a region
in Figure 3b (lower spectrum) overlaid with the best-fit scale (solid
curve) and absorption (dotted curve) models.  This example illustrates
how a negative $\Delta N_{\rm H}$ enhances the lower energy portion of
the spectrum.  However, neither model provides a good fit because
oxygen emission, i.e., the lines at 0.53 and 0.65 keV, appears
relatively enhanced in this particular spectrum. Large positive column
densities are found in interior regions of the remnant, especially in
the vicinity of the triangular hole mentioned previously.  We plot an
example spectrum in Figure 3c (lower spectrum) where the absorption
model clearly provides a better fit than the scale model.  Much of the
rest of the remnant shows $F$-values with low significance ($<
1\sigma$) that correspond to $\Delta N_{\rm H}$ near zero.  These
represent regions of mean spectra, as illustrated in Figure 3a (lower
spectrum).

\subsection{Spectral Analysis}

To confirm that the image analysis produced valid and numerically
accurate results, we carried out spectral fits of nonequilibrium
ionization (NEI) thermal plasma models, using the planar shock model
described in \citet{hughes00}. We summed together a number of
individual regions from the image analysis to obtain three
well-sampled spectra, each containing at least 50,000 counts,
corresponding to the soft, hard, and mean spectral regions identified
above.  The extraction region for the soft spectrum came from along
the southeastern limb of the remnant, approximately centered at
$5^{\rm h}35^{\rm m}42^{\rm s}, -66^\circ02'07''$.  The hard spectrum
came from a region centered on $5^{\rm h}35^{\rm m}47^{\rm s},
-66^\circ02'35''$ and encompassing the darkest portions of the
grayscale image of Figure 2c. The mean spectrum was extracted from
pixels, distributed throughout the interior of the SNR, that had low
significant ($< 1\sigma$) $F$-values.  These spectra are also plotted
in Figure 3 along with their best fit NEI models (upper spectra).  We
allowed the temperature, terminal value of the ionization timescale,
column density, and the elemental abundances for each spectrum to be
free parameters.  Again, the column density was determined using LMC
abundances.

The best-fitted parameter values are quoted in Table 1. Abundances are
lower than solar and generally consistent with the gas phase LMC
values \citep{rusdop92}.  The abundances are slightly elevated near
the absorbed region, which can be seen in the form of somewhat
stronger Ne and Mg line intensities (compared to the mean spectrum) in
Figure 3c.  This result is less secure than the formal statistical
errors would indicate, due to uncertainties in modelling the Fe
L-shell emission, and therefore we do not discuss this result further.
Also, there is a slight excess over the thermal model fits in the mean
and especially hard spectra.  Although the statistics are low, the
3.3--7.0 keV band image, where this excess appears, generally
resembles the rest of the X-ray remnant; there is no particular
enhancement at the site of the brightest radio or optical
emission. The spectral fits for the integrated soft rim spectrum yield
a lower column density and higher $kT$ than the mean.  We discuss
these fits in more detail below.

To compare the spectral-fit and image-analysis results, we had to
average the values of $\Delta N_{\rm H}$ from the image analysis over
the larger regions used for the spectral study.  This was done by
taking an average, weighted by the number of counts, of the individual
$\Delta N_{\rm H}$ derived from each separate adaptive mesh pixel.
These average values of $\Delta N_{\rm H}$ (see Table 2) agree fairly
well with the $\Delta N_{\rm H}$ calculated from the difference
between the soft or hard spectral fit $N_{\rm H}$ and the value from
the mean spectrum.  This test validates the image analysis results and
gives us confidence that the hard and soft regions identified in
Figure 2 do indeed differ significantly from the mean.

\citet{levenson} estimated the interstellar reddening toward N63A to
be $E_{B-V}$ = 0.25, using the Balmer decrement. From established
correlations between $N_{\rm H}$ and optical extinction
\citep{gorenstein,predehl} and assuming $A_V \approx 3 E_{B-V}$
\citep{gorenstein}, we obtain $N_{\rm H}$ = (1.3--1.7) $\times 10^{21}$
cm$^{-2}$.  This agrees nicely with our value for the mean $N_{\rm H}=(1.68
\pm 0.12) \times10^{21}$ cm$^{-2}$ and provides further confirmation
of our spectral fits.

\subsection{Limits on a Compact Object}

As mentioned previously, we find no evidence from broadband \chandra\
images for a hard unresolved X-ray source in N63A. Our image analysis
technique, however, allows us to search the entire remnant for a
region (or regions) that are dominated by any specific spectral form.
To simulate the X-ray emission from a young PWN we used a power-law
spectral model with photon index $\alpha_p = 2$ and again allowed the
additional hydrogen column density to be free.  None of the 570
regions provided an acceptable fit (the minimum reduced-$\chi^2$ was
6.67), indicating that nowhere within N63A is the emission dominated
by a hard power-law spectrum.  We determined a numerical upper limit
to the flux of a hard source using an image from the 3.25--6.40 keV
band, which is free of strong emission lines in the integrated
spectrum.  The 3$\sigma$ count rate upper limit for an unresolved
source anywhere in the interior of N63A is $<$ $5.4\times 10^{-4}$
s$^{-1}$ or an unabsorbed 2--8 keV flux of $2.5\times 10^{-14}$ ergs
cm$^{-2}$ s$^{-1}$.  The limit for a source with a diameter of 2 pc,
roughly the size of a young PWN, is $<$ $3.0\times 10^{-3}$ s$^{-1}$,
implying an unabsorbed 2--8 keV flux of $1.4\times 10^{-13}$ ergs
cm$^{-2}$ s$^{-1}$ or luminosity of $4\times 10^{34}$ ergs s$^{-1}$.
This limit allows us to rule out the presence of a bright PWN powered
by a young, energetic pulsar like the Crab, or the two already known
in the LMC (N157B and 0540$-$69).  However we cannot exclude an older
PWN, like the one in Vela ($\sim$11,000 yr old), or one powered by an
intrinsically weaker pulsar, such as the recently discovered PWN in
the much younger ($\sim$1600 yr) SNR G292.0+1.8 \citep{hughes01}.

\section{DISCUSSION}

The crescent-shaped plumes in N63A resemble features seen in the {\em
ROSAT} image of the Vela supernova remnant \citep{asch}.  These are
the only SNRs where such features have been detected.  The plumes in
N63A extend from the main shock boundary, distorting the otherwise
very symmetrical shape of the remnant.  The largest one in the
southwest reaches approximately 4.2 pc (in projection) beyond the main
shock wave, the one just north of this protrudes about 3.5 pc, and the
smaller northern region about 1.6 pc.  The lateral size of the largest
crescent feature is about 6 pc measured halfway out from the shell.
These sizes should be compared to the 8.2 pc radius of the nearly
circular main blast wave of N63A.  Three other crescent features
appear closer in projection to the main shell in the northeast, east,
and south.  \citet{wang} use two-dimensional hydrodynamic simulations
to model the formation of such structures.  They find that dense, high
velocity clumps of ejecta may protrude beyond the forward shock as the
remnant interacts with the surrounding medium.  Shocks moving through
the clump at first crush it and then cause it to expand laterally,
eventually taking on the shape of a crescent, such as is seen in N63A.

The origin of crescent-shaped features as ejecta is supported by
\citet{miyata}, who found an overabundance of Si in their study of
Vela shrapnel A.  {\em ASCA} and \chandra\ observations of Vela bullet
D show strong O, Ne, and Mg emission \citep{slane,plucinsky},
suggesting an origin as a dense knot of supernova ejecta.  However,
\citet{plucinsky} argue that a nearly-solar abundance plasma far from
ionization equilibrium is a more likely explanation for their
observations. N63A is about 200 times more distant than Vela, limiting
our ability to do a similar detailed spectral study. We find that none
of the crescent regions in N63A shows strongly enhanced abundances,
although there is weak evidence for enhanced Ne, Mg, and Si line
emission at the apex of the northern crescent (compared to the sides).
We can confidently conclude that the plumes are not {\em dominated} by
ejecta.  Thus, for the high-speed ejecta clump scenario to be correct,
the clumps in N63A must be significantly mixed with ambient material,
perhaps signalling the onset of their fragmentation and destruction.

On the other hand, our image analysis has revealed that both the
crescent-shaped regions and the entire rim of the SNR show
considerably softer spectra than average for N63A.  This argues for a
common origin in terms of shocked ISM.  However, confirmation of this
hypothesis will require more careful study of spatially resolved X-ray
spectra from these regions. Based on models of SNR evolution as well
as other \chandra\ observations (e.g. Rakowski et al.~2002), we expect
the most rapid changes in the thermodynamic state of the post-shock
gas to occur immediately behind the blast wave, and it is at the
projected rim where these variations can be most cleanly resolved.  It
is somewhat puzzling, therefore, that our initial fits for the
integrated soft rim spectrum appear to support a relatively lower
column density, which, if taken literally, would imply a screen of
absorbing matter covering most of the SNR that falls off near the
projected rim.  We find this to be a rather contrived geometry and
believe the low fitted value of $N_{\rm H}$, as well as the higher
fitted value of $kT$, to be spurious and a result of our
oversimplified spectral modelling, i.e., the use of a single component
NEI planar shock model for a region with spatially varying spectral
characteristics.  In fact when the rim region is examined in finer
detail, it is apparent that there is spectral evolution of the
ionization state with position behind the shock front on scales of an
arcsec or so, qualitatively showing the advance of post-shock
ionization state. Further work is in progress to study this effect and
use it to quantify the dynamical state of the remnant.

We interpret the hard region to be absorption by an intervening cloud
in the LMC with a mass of $\sim$450 $M_\odot$, determined by
integrating the $N_{\rm H}$ map over the extraction region for the
spectrum in Figure 3c.  If we assume that the cloud's depth along the
line-of-sight is roughly half of its length (3.8 pc), which is
reasonable given our discussion below, then the mean density of the
cloud is on the order of 250 cm$^{-3}$. This value is consistent with
estimates of $\sim$50--300 cm$^{-3}$ (see Shull 1983 and references
cited therein) for the preshock density of the optical remnant.  Two
other smaller absorbing clouds, north and northeast of center, are
roughly 20-30 $M_\odot$ in size.  The large region of excess
absorption is situated in nearly the same area as the bright radio and
optical emission from N63A (see Figure 4).  To explain the X-ray
absorption, the cloud must be on the near side of N63A with little to
no residual X-ray emission coming from in front of it.  The radio and
optical emission signal the presence of slow shocks (several 100 km
s$^{-1}$ or less) driven into the dense gas of the cloud.  However,
the modest extinction to the optical knots (consistent with the mean
$N_{\rm H}$ to N63A rather than the higher value associated with the
cloud) means that we are predominantly seeing the shocks being driven
into the {\em near} side of the cloud. This is further supported by
the observation of only blue-shifted emission from the shock-heated
optical lobes \citep{shull}.  In essence, then, we are viewing the
dense cloud as it is being engulfed by the remnant's blast wave.  The
clumpy, knot-like features apparent in the absorption map and {\em
HST} image suggest that the dense cloud's interaction with the blast
wave is causing it to break-up and fragment. Indeed the highest
measured radial velocities in the optical band (up to 250 km s$^{-1}$)
appear mainly around the edges of the lobes \citep{shull}, while the
interior portions appear to be moving toward us at less than half this
speed. Numerical simulations of an ISM cloud overrun by a SN
blast-wave (e.g., Klein, McKee, \& Colella 1994) show, rather
generically, that the edges of the cloud are accelerated more rapidly
(hence attaining higher speeds) than the cloud core. When viewed
perpendicular to the direction of motion of the blast wave, this
results in a characteristic shape for the shocked cloud: the edges are
swept forward along with the blast wave, while the core trails behind.
An example of this type of interaction can be seen in an isolated
cloud on the eastern limb of the Cygnus Loop \citep{fesen}. If the
shocked cloud is viewed {\em along} the direction of the blast wave's
motion, then the edges of the cloud should show higher radial
velocities than the core.  This situation appears, at least
qualitatively, to describe the optical characteristics of
N63A. Further studies of this remnant across the wavebands should help
to confirm this scenario and provide deeper insights into the
interaction of SNRs with their environments.

\acknowledgements 

We would like to thank Cara Rakowski for help on numerous \chandra\
issues as well as for allowing us to use her mesh program. We also
thank John Dickel for sharing his 6-cm radio image with us, Parviz
Ghavamian for helpful discussions and comments on the text, and John
Nousek and Dave Burrows for assistance with the original
proposal. This research was partially supported by \chandra\ grant
GO0-1035X (JPH) and NASA contract NAS8-39073 (POS).

\clearpage

\begin{deluxetable}{cccc}
\tablewidth{0pt}
\tablecaption{Best-Fit Spectral Parameters}
\tablehead{\colhead{~} & \colhead{Mean} & \colhead{Soft} & \colhead{Hard}}
\startdata
$N_{\rm H}$ ($10^{21}$ cm$^{-2}$)&$1.68 \pm 0.12$&$0.86 \pm 0.15$&$3.03 \pm 0.14$\\
$kT$ (keV)&$0.78 \pm 0.02$&$0.96 \pm 0.08$&$0.76 \pm 0.01$\\
$\log(n_et/{\rm cm^{-3} s})$&$11.66 \pm 0.16$&$11.15 \pm 0.09$&$12.14 \pm 0.11$\\
O  & $0.18 \pm 0.06$ & $0.21 \pm 0.02$ & $0.39 \pm 0.06$\\
Ne & $0.26 \pm 0.04$ & $0.39 \pm 0.04$ & $0.65 \pm 0.09$\\
Mg & $0.16 \pm 0.02$ & $0.28 \pm 0.05$ & $0.43 \pm 0.05$\\
Si & $0.15 \pm 0.02$ & $0.19 \pm 0.04$ & $0.22 \pm 0.03$\\
S  & $0.32 \pm 0.07$ & $0.15 \pm 0.14$ & $0.43 \pm 0.08$\\
Fe & $0.13 \pm 0.01$ & $0.17 \pm 0.03$ & $0.10 \pm 0.01$\\
$\chi^2/d.o.f.$&287.51/163&256.20/126&328.93/175\\
\enddata
\tablecomments{Quoted errors are for a 90\% confidence level.}
\end{deluxetable}

\clearpage

\begin{deluxetable}{ccc}
\tablewidth{0pt}
\tablecaption{$\Delta N_{\rm H}$ Values}
\tablehead{\colhead{~} & \colhead{Soft} & \colhead{Hard}}
\startdata
$\Delta N_{\rm H}$ - spectral fit\\($10^{21}$ cm$^{-2}$)&$-0.82 \pm 0.27$&$1.35 \pm 0.25$\\
$\Delta N_{\rm H}$ - image analysis\\($10^{21}$ cm$^{-2}$)&$-0.94$&$2.20$\\
\enddata
\tablecomments{$\Delta N_{\rm H}$ refers to the difference between the soft or hard $N_{\rm H}$ values and the mean value.}
\end{deluxetable}

\clearpage

\figcaption{(a) Three-color X-ray image of N63A.  The colors are
reversed to better illustrate the structure.  Cyan (blue color)
represents the 0.16--0.75 keV energy band; magenta (pink color)
represents the 0.75--0.96 keV band, and yellow represents the
0.96--6.64 keV band.  North is up and east is to the left.  (b) Color
composite figure of N63A composed from 8638 MHz ATCA radio (red), {\em
HST} H$\alpha$ optical (green) and broadband \chandra\ X-ray (blue)
images. North is up and east is to the left.}

\figcaption{(a) {\em Chandra} broadband X-ray image (0.4--7 keV) of
N63A.  North is up and east is to the left.  (b) Regions of soft
spectra overlaid with X-ray contours.  Darker regions indicate more
negative $\Delta N_{\rm H}$, the lowest of which is $-1.9 \times
10^{21}$ cm$^{-2}$.  (c) Regions of hard spectra overlaid with X-ray
contours.  Darker regions indicate higher $\Delta N_{\rm H}$, the
largest of which is $7.4 \times 10^{21}$ cm$^{-2}$.}

\figcaption{(a) Mean spectra: \textit{upper} - Summed spectra of regions
with zero $\Delta N_{\rm H}$ totalling greater than 50,000 counts,
along with the best-fit spectral model, \textit{lower} - Spectrum of
one region with zero $\Delta N_{\rm H}$, along with best-fit scale
model.  (b) Soft spectra: \textit{upper} - Summed spectra of regions
with large negative $\Delta N_{\rm H}$ containing at least 50,000
counts, along with the best-fit spectral model and mean spectrum,
\textit{lower} - Spectrum of one region with large negative $\Delta
N_{\rm H}$, along with best-fit absorption (dotted) and scale (solid)
models.  (c) Hard spectra: \textit{upper} - Summed spectra of regions
with large positive $\Delta N_{\rm H}$ containing at least 50,000
counts, along with the best-fit spectral model and mean spectrum,
\textit{lower} - Spectrum of one region with large positive $\Delta
N_{\rm H}$, along with best-fit absorption (dotted) and scale (solid)
models.}

\figcaption{Contours of excess column density (solid curves) from
Figure 2c after smoothing with a gaussian function ($\sigma =
1^{\prime\prime}$) superposed on a grayscale image of the {\em HST}
[\ion{S}{2}] 6732\AA~WFPC2 data \citep{chu1}.  The solid contours are
plotted at values of $\Delta N_{\rm H} = 0.75$, 1.5, 3, and 6 $\times
10^{21}$ cm$^{-2}$.  Dotted contour values correspond to the broadband
{\em Chandra} X-ray image and are the same as in Figure 2.  North is
up and east is to the left.}

%\end{document}

\clearpage

\plotone{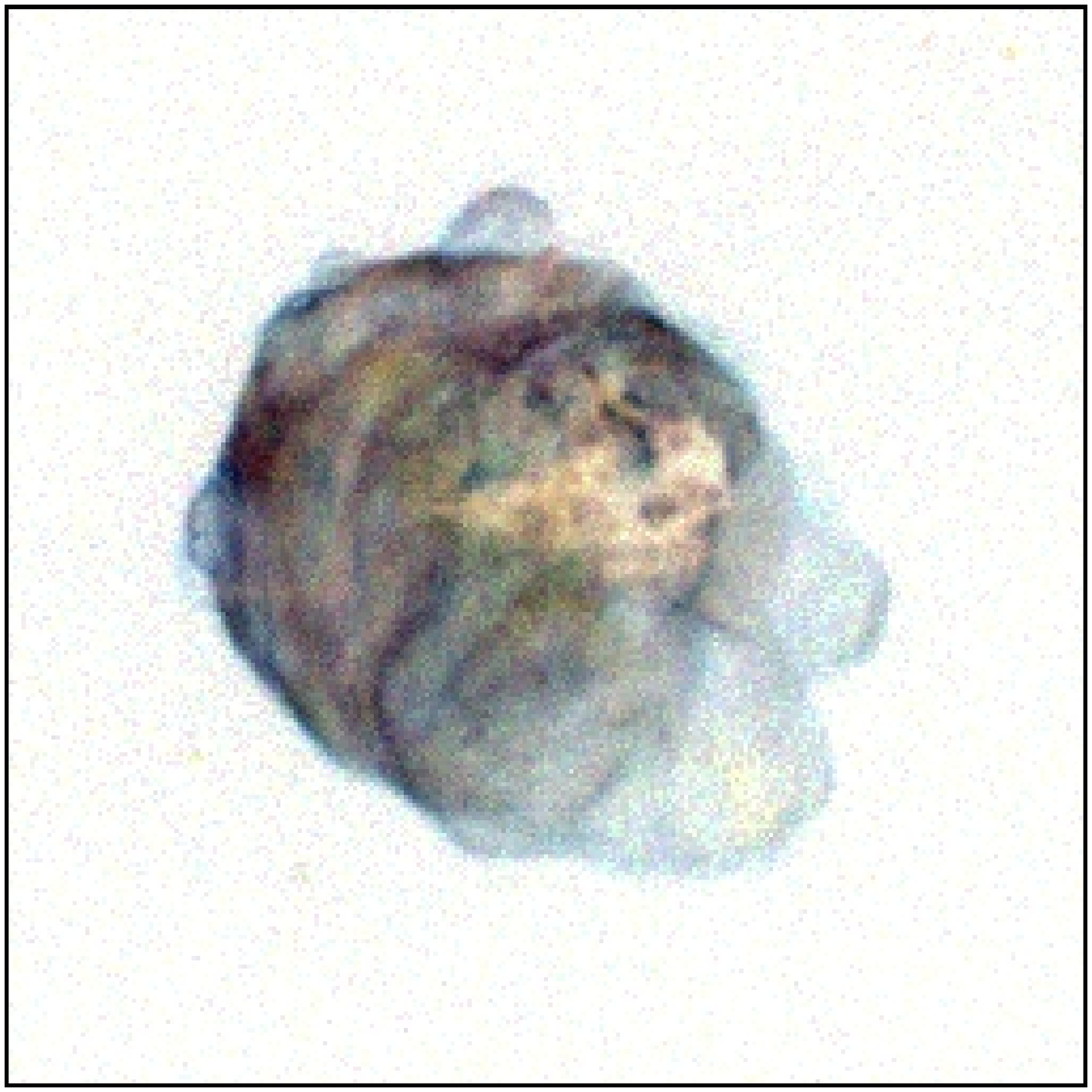}

\clearpage

\plotone{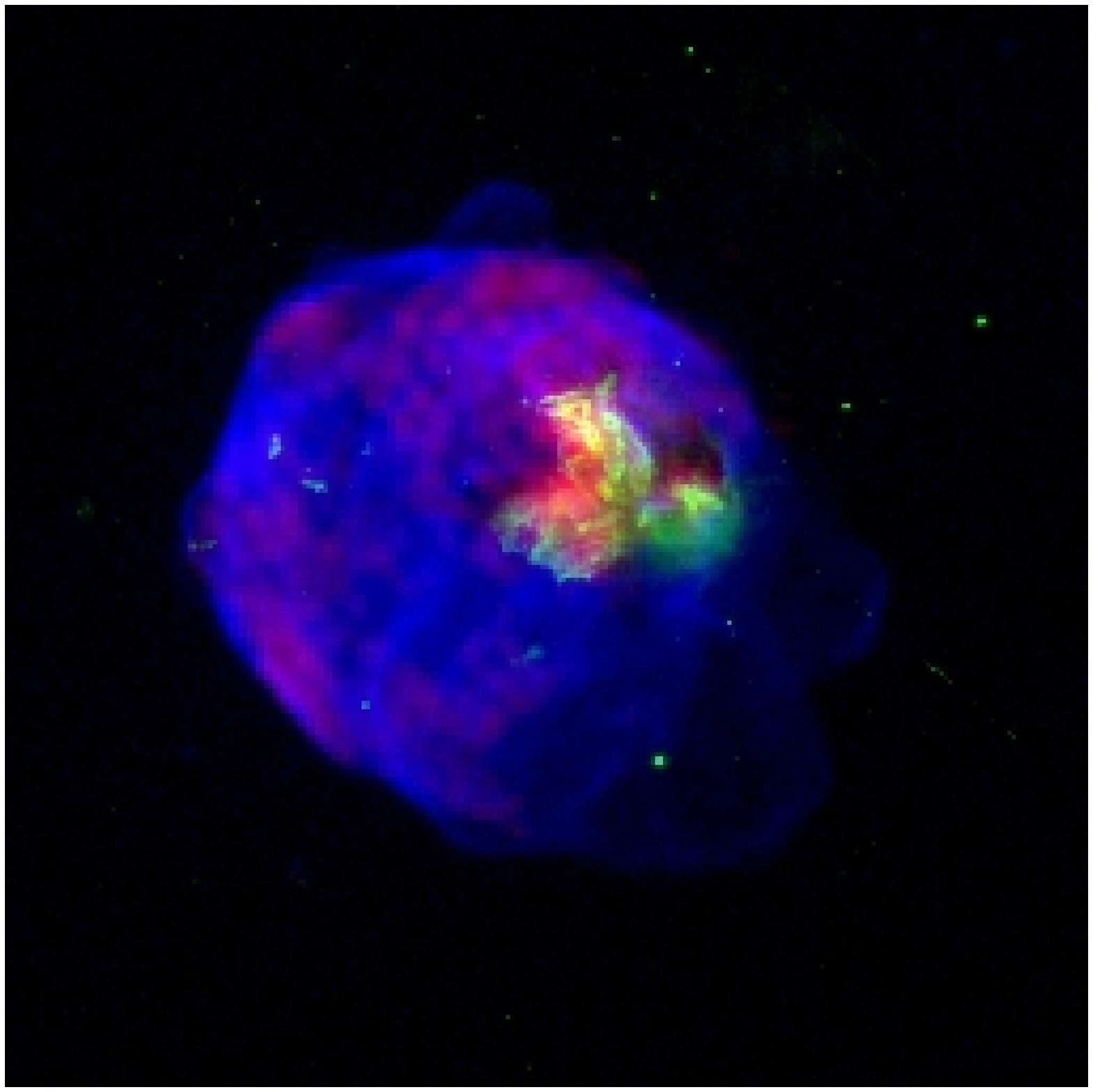}

\clearpage

\plotone{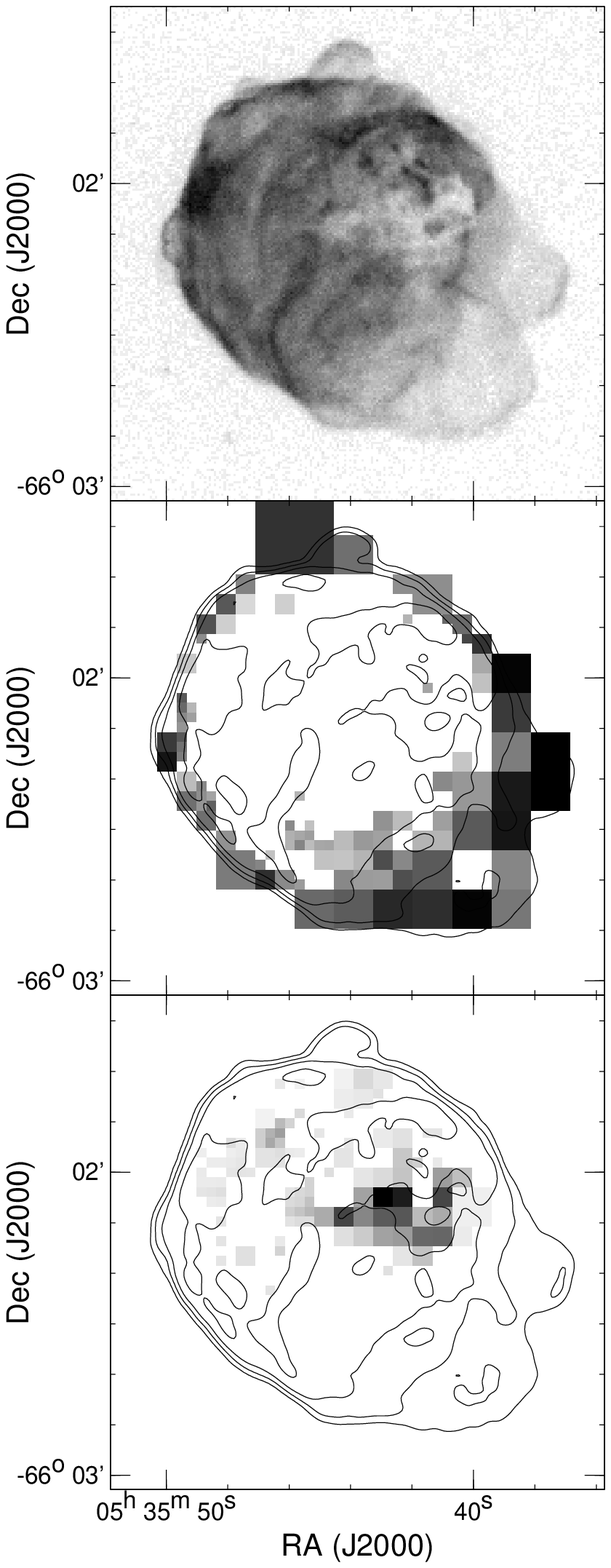}

\clearpage

\plotone{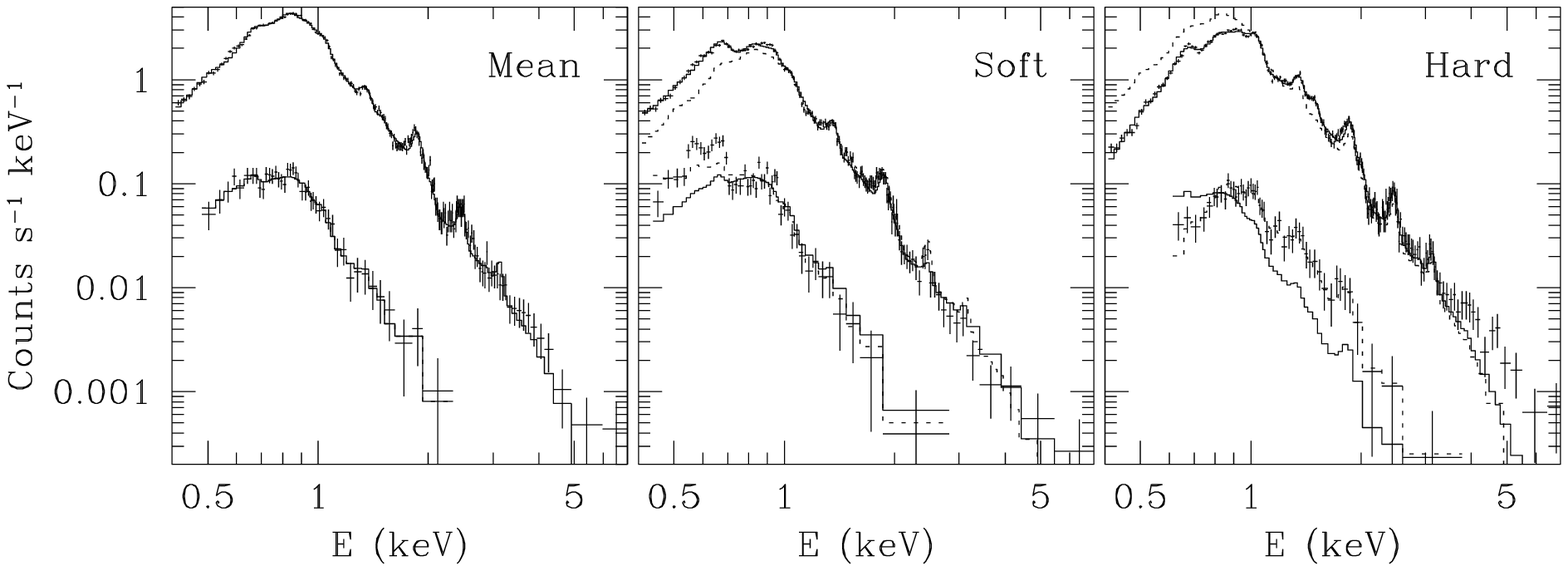}

\clearpage

\plotone{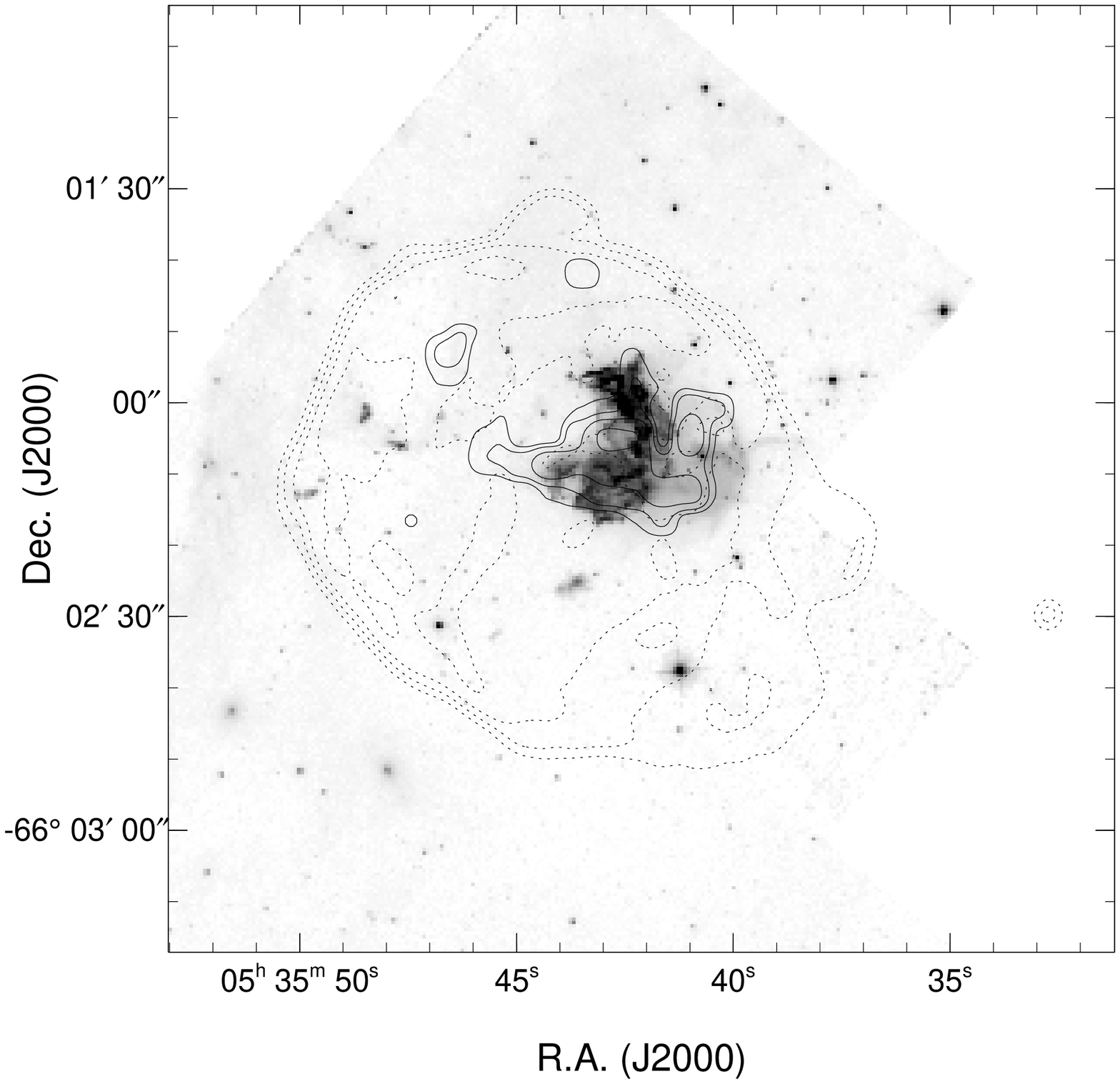}

\end{document}